\documentclass[prd,preprint,tightenlines,floatfix,showpacs,preprintnumbers,nofootinbib,eqsecnum]{revtex4}
\pdfoutput=1

 \usepackage[dvips,final]{graphicx}
  \usepackage{amssymb}	
   \usepackage{amsmath}
    \usepackage{amsfonts}
     \usepackage{epsfig}
      \usepackage{bm} % bold math
      \usepackage[dvipsnames,usenames]{color}
      \usepackage{comment}
      \usepackage{float}
      
      \usepackage[utf8]{inputenc}

\begin{document}

\title{Probing double parton scattering via associated open charm and bottom production
in ultraperipheral $pA$ collisions}

\author{Edgar Huayra$^{1}$}
\email{yuberth022@gmail.com}

\author{Emmanuel G. de Oliveira$^{1}$}
\email{emmanuel.de.oliveira@ufsc.br}

\author{Roman Pasechnik$^{1,2,3}$}
\email{Roman.Pasechnik@thep.lu.se}

\affiliation{
\\
{$^1$\sl Departamento de F\'isica, CFM, Universidade Federal 
de Santa Catarina, C.P. 476, CEP 88.040-900, Florian\'opolis, 
SC, Brazil
}\\
{$^2$\sl
Department of Astronomy and Theoretical Physics, Lund
University, SE-223 62 Lund, Sweden
}\\
{$^3$\sl Nuclear Physics Institute ASCR, 25068 \v{R}e\v{z}, 
Czech Republic\vspace{1.0cm}
}}

%%%%%%%%%%%%
\begin{abstract}
\vspace{0.5cm}
In this article, we propose a novel channel for phenomenological studies of the double-parton scattering (DPS) 
based upon associated production of charm $c\bar{c}$ and bottom $b\bar{b}$ quark pairs in well-separated rapidity 
intervals in ultra-peripheral high-energy proton-nucleus collisions. This process provides a direct access 
to the double-gluon distribution in the proton at small-$x$ and enables one to test the factorised DPS pocket 
formula. We have made the corresponding theoretical predictions for the DPS contribution to this process at 
typical LHC energies and beyond and we compute the energy-independent (but photon momentum fraction 
dependent) effective cross section.
\end{abstract}
%%%%%%%%%%%%


\maketitle

%%%%%%%%%%%%%%%%%%%%%%%%
\section{Introduction}
\label{Sect:intro}
%%%%%%%%%%%%%%%%%%%%%%%%

With an increase of collision energy, the probability for more than one parton-parton scattering to occur 
in the same proton-proton or proton-nucleus collision grows faster compared to that of the single-parton 
scattering (SPS) leading to the well-known phenomenon of multi-parton interactions (MPIs) known since a long 
time ago \cite{Paver1982,Mekhfi:1983az,PhysRevD.36.2019}. Due to measurements at the Large Hadron Collider (LHC), 
the physics of MPIs has attracted a lot of attention from both theoretical and experimental communities 
(for recent works on this topic, see e.g.~Refs.~\cite{Gaunt:2009re,Diehl:2011yj,PhysRevD.85.114009,Aaij:2015wpa,
Blok:2016lmd,Rinaldi:2015cya,PhysRevD.97.094010,PhysRevD.47.4857,Aaboud:2018tiq} and references therein). A first non-trivial 
example, the double-parton scattering (DPS), becomes particularly significant in production of specific 
multi-particle final states such as meson pairs \cite{PhysRevD.97.094010}, four identified jets \cite{PhysRevD.47.4857} 
or leptons \cite{Aaboud:2018tiq} etc. These processes are traditionally considered as an important source 
of information about a new class of non-perturbative QCD objects, the double-parton distribution functions 
(dPDFs) being now actively explored in the literature. They describe the number density and correlations 
of two colored partons in the proton, with given longitudinal momentum fractions $x_1$, $x_2$ and placed 
at a given transverse relative separation ${\bf b}$ of the two hard collisions \cite{PhysRevD.60.054023} 
(for a detailed review on theoretical grounds, see e.g.~Ref.~\cite{Diehl:2017wew} and references therein).

While complete theoretical predictions for dPDFs involving the unknown nonperturbative QCD parton correlation 
functions are not available, a few model calculations exist attempting to pick the most significant features of dPDFs \cite{PhysRevD.87.034009,PhysRevD.87.114021,Rinaldi:2014ddl,Rinaldi:2015cya,Blok:2016lmd}. In order to perform 
any comprehensive verification of such models, much more 
phenomenological information is needed as no direct measurement or extraction of dPDFs from the experimental 
data has yet been possible. Experimentally, a distinctive signature of DPS associated with the so-called 
effective cross section, $\sigma_{\rm eff}$, has already been identified and measured in different channels 
at central rapidities (see e.g.~Refs.~\cite{Akesson1987163,ALITTI1991145,PhysRevD.47.4857,Aaij2012,
PhysRevD.81.052012,Aad:2014kba,Aaij:2016bqq,Sirunyan:2017hlu,Aaboud:2018tiq}), while many Monte-Carlo 
generators naturally incorporate MPIs as part of their framework \cite{PhysRevD.36.2019}.

The effective cross section is experimentally defined as ratio of double to product of two single inclusive 
production rates for final-state $A_1$ and $A_2$ systems in two independent hard scatterings and 
represents the effective transverse overlap area containing the interacting hard partons. 
With this definition, the DPS cross section is estimated as~\cite{DIEHL2011389,Diehl:2011yj,PhysRevD.85.114009} 
(for a detailed review on this topic, see e.g.~Refs.~\cite{Bartalini:2011jp,Bansal:2014paa}),
\begin{eqnarray}
\sigma^{A_1A_2}_{\rm DPS} = \frac{\kappa}{2} \frac{\sigma^{A_1}_{\rm SPS}\sigma^{A_2}_{\rm SPS}}{\sigma_{\rm eff}} \,,
\label{eq:pocket}
\end{eqnarray}
where $\sigma^{A_{1,2}}_{\rm SPS}$ represents the corresponding SPS cross section for 
production of $A_{1,2}$ systems, and $\kappa$ is the symmetry factor 
depending on whether the final states are the same ($A_1 = A_2$, $\kappa = 1$) 
or different ($A_1 \not= A_2$, $\kappa = 2$). In general, $\sigma_\text{eff}$ depends on scales, momentum fractions, and parton flavours involved. In many theoretical studies it is however assumed that it is a constant geometrical factor; under this approximation (\ref{eq:pocket}) is known as the ``pocket formula''.

Among the hadron final states, double open heavy flavor production is considered to be an important and promising tool 
for probing the DPS mechanism \cite{PhysRevD.85.094034}. In particular, the LHCb Collaboration has recently reported 
an enhancement in the data on double charm production cross section in $pp$ collisions \cite{Aaij2012,Aaij2014} that 
could not be described without a significant DPS contribution as was found in Ref.~\cite{PhysRevD.87.074039}. 
More possibilities have been recently discussed also in the case of $c\bar c b \bar b$ and $b\bar b b \bar b$ 
final states, as well as in associated production of open heavy flavor and jets, in Refs.~\cite{PhysRevD.66.074012,PhysRevD.88.034005,PhysRevD.97.094010,PhysRevD.96.074013}.

Within yet large experimental uncertainties, the c.m.\ collision energy dependence of the effective cross section 
is consistent with a constant $\sigma_{\rm eff}\sim 15-20$ mb for the channels probed by most of the existing measurements \cite{Akesson1987163,ALITTI1991145,PhysRevD.47.4857,Aaij2012,PhysRevD.81.052012,Aad:2014kba,Aaij:2016bqq,
Sirunyan:2017hlu,Aaboud:2018tiq}. However, in associated production of heavy quarkonia such as double-$J/\psi$ 
and $J/\psi \Upsilon$, one discovers systematically lower values of $\sigma_{\rm eff}$ than in all the other channels 
studied so far \cite{Abazov:2014qba,Abazov:2015fbl,Aaboud:2016fzt,Khachatryan:2016ydm}. Such a discrepancy may hint towards 
a non-universality of $\sigma_{\rm eff}$ due to e.g. spatial fluctuations of the parton densities \cite{Mantysaari:2016ykx}. 
Typically, measurements of the DPS contributions for different production processes 
need a dedicated experimental analysis and tools, and the precision is usually very limited and suffers due to large 
backgrounds coming from the standard SPS processes. 

The use of ultra-peripheral $pA$ collisions (UPCs) for probing the DPS mechanism and further constraining the effective cross section has not yet been properly studied in the literature. In contrast, the SPS UPC case has been studied in, e.g., Refs.~\cite{Klein:2002wm, Adeluyi:2012sw}. In UPCs, the high-energy colliding systems pass each other at large transverse separations and thus do not undergo hadronic interactions. In this case, they interact electromagnetically via an exchange of quasi-real photons. The corresponding Weisz\"acker-Williams (WW) photon flux
\cite{vonWeizsacker:1934nji,Williams:1934ad} is scaled with the square of electric charge of the emitter and is thus strongly enhanced for a heavy nucleus making the $pA$ and $AA$ UPCs more advantageous compared to that in $pp$ collisions. It is worth noticing that the photon spectrum of a heavy nucleus is rather broad, where the peak-energy in the target rest frame scales linearly with the nuclear Lorentz factor which represents yet another advantage of UPCs. Finally, an additional reduction of the backgrounds is provided by tagging on the final-state nucleus identifying the momentum transfer taken by the exchanged photon, together with reconstructing the four-momenta of the produced final-state particles.
%=======================================================
\begin{figure*}[htb]
 \begin{center}
 {\includegraphics[width=0.7\textwidth]{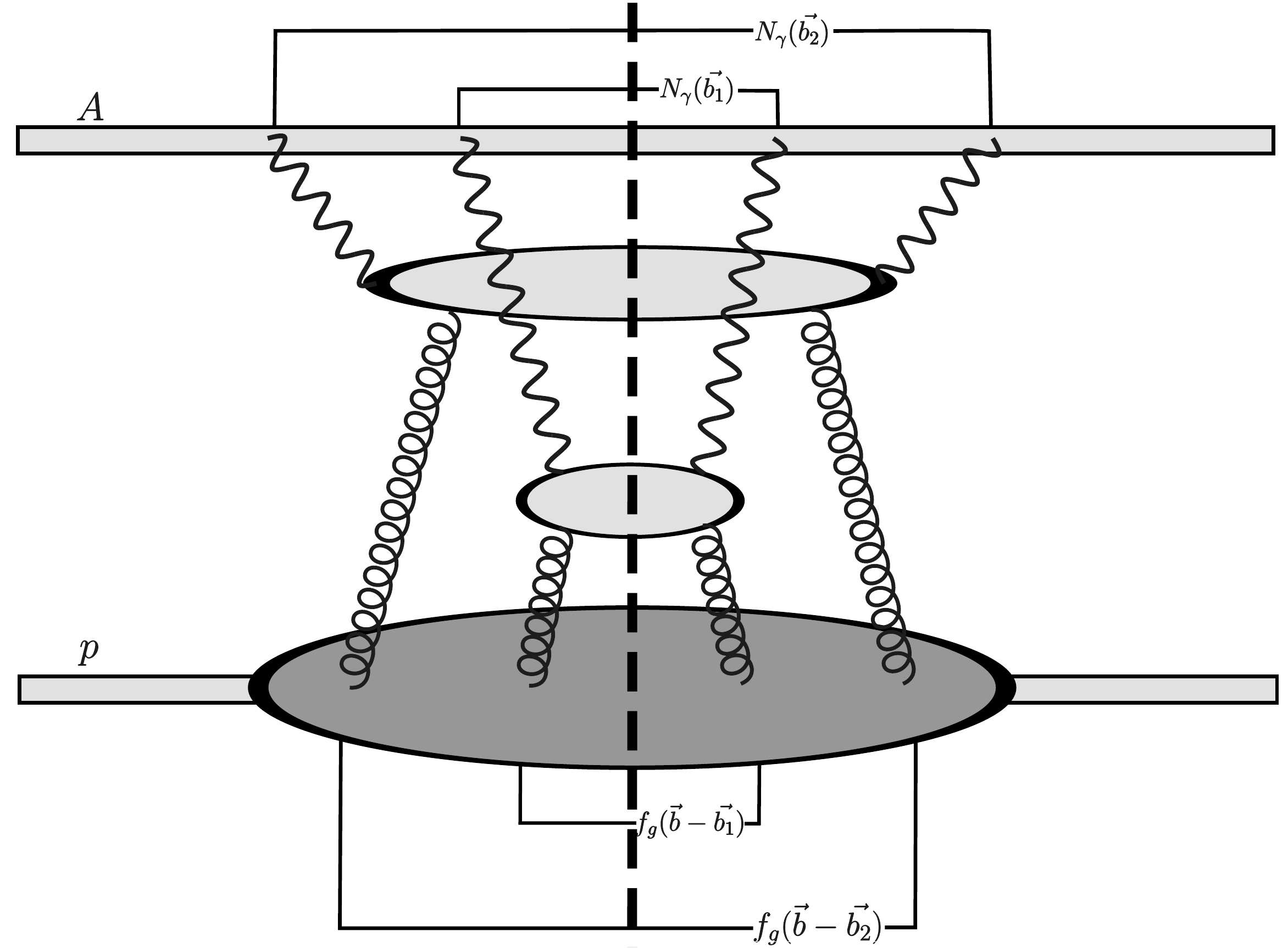}}
 \end{center}
\caption{
A schematic illustration of the $A+p \to A + (c\bar c b \bar b) + X$ cross section in $pA$ UPCs.
}
\label{fig:DPS-scheme}
\end{figure*}
%=======================================================

In this work, we explore possibilities for a new measurement of the gluon dPDF in the proton at small-$x$ by means 
of $A+p \to A + (c\bar c b \bar b) + X$ reaction in high-energy $pA$ UPCs schematically illustrated 
in Fig.~\ref{fig:DPS-scheme}. This process offers interesting possibilities and a cleaner environment 
for probing the DPS contribution compared to that in $pp$ collisions. 

In order to study the corresponding reaction in UPCs, we have to compute an effective cross section 
for the interaction of two partons (e.g. gluons) on one side with two photons on the other side, in contrast 
to the four parton case in regular $pp$ or $pA$ collisions. At small-$x$ and for hadronic final states, these partons are overwhelmingly likely to be gluons, so in the subsequent discussion we will simply refer to them as gluons. By exchanging the two-gluon initial state by two photons on the nucleus side, the effective cross section is expected to increase significantly. 
This is due to the fact that two photons from a single nucleus overlap much less than the two gluons, 
since the latter are well localized inside the nucleons, while photons are more spread out, 
specially in the case of UPCs, when they are required to be outside the nucleus. 
As far as we know, this effective cross section has not yet been calculated or measured earlier, 
but it is clearly important in order to better understand the impact parameter dependence of 
parton distributions in general.

Regarding the order in coupling constants, the DPS process is of the order of $(\alpha \alpha_s)^2$, 
while the SPS process -- $\alpha \alpha_s^3$, i.e. the SPS cross section is by default a factor of 
$\alpha_s/\alpha$ larger. However, the DPS reaction is expected to dominate the $c\bar c b \bar b$ production 
cross section over the SPS one at high energies, particularly, for a large separation between rapidities of $c\bar c$ 
and $b \bar b$ pairs. Indeed, in the case of a large invariant mass of the $c\bar c b \bar b$ system, 
the parton distributions are computed at larger $x$ for the SPS case than that in DPS, since more energy 
in the initial state is needed, especially if there is a considerably large rapidity difference between 
the $c\bar c$ and $b\bar b$ pairs. As the PDFs decrease very fast with $x$ in the case of gluons 
(and photons likewise), the SPS process is expected to be suppressed.

Thus, in order to extract the DPS contribution to this process, one should 
consider light $c\bar c$ and $b \bar b$ pairs produced at relatively large rapidity separation 
$\delta Y=Y_{c\bar c} - Y_{b\bar b} \gg 1$. This is required in order to maximize the invariant 
mass of the SPS $\gamma+g \to c\bar c b \bar b$ background process, and hence to sufficiently suppress the background compared to the DPS contribution whose dependence on $\delta Y$ is expected to be flatter. 
In the case of $c\bar c c \bar c$ and $b\bar b b \bar b$ production, however, such a separation 
would be much more difficult (if not impossible) since combining a quark $Q$ and antiquark $\bar Q$ 
of the same flavor does not guarantee that they come from the same SPS process $\gamma+g\to Q \bar Q$. 
The relative $\delta P_\perp = P_\perp^{c\bar c} - P_\perp^{b\bar b}$ variable is of less importance 
for the SPS background suppression since both the SPS and DPS components are peaked around 
a small $\delta P_\perp\approx 0$, while at large $\delta P_\perp$ the DPS term is nonzero 
only at the NLO level. 

The relative SPS background suppression at large $\delta Y$ is only a qualitative expectation based upon simple kinematical arguments mentioned above. In this work, however, we are focused only on the DPS contribution to the $c\bar c$ and $b \bar b$ pairs production. The detailed analysis of the SPS $\gamma+g \to c\bar c b \bar b$ amplitude and the corresponding differential cross section falls beyond the scope of the current work and will be performed elsewhere.

The paper is organised as follows. In Sec.~\ref{Sect:ccbb-DPS} we derive the formula for the UPC double heavy quark photoproduction, that is written with the help of an effective cross section that depends on the photon longitudinal momentum fraction. We also review the key components of such calculation. In Sec.~\ref{Sect:Results} we present our numerical results at LHC and larger energies. We conclude our paper in Sec.~\ref{Sect:Concl}.

%%%%%%%%%%%%%%%%%%%%%%%%%%%%%%%%%%%%%%%%%
\section{Double quark-pair production in UPC: DPS mechanism}
\label{Sect:ccbb-DPS}
%%%%%%%%%%%%%%%%%%%%%%%%%%%%%%%%%%%%%%%%%

In the high-energy limit, the cross section for $c\bar{c}b\bar{b}$ production via DPS can be represented as 
a convolution of the impact-parameter dependent differential probabilities to produce separate 
$c\bar{c}$ and $b\bar{b}$ pairs in $pA$ collisions
\begin{eqnarray}
\frac{d^4\sigma_{pA \rightarrow XA + c\bar{c} + b\bar{b}}}{dy_{c}dy_{\bar{c}}dy_{b}dy_{\bar{b}}}  =  
\int d^2  \vec{b} \, \Theta(b - R_A - R_p) \frac{d^2P_{pA \rightarrow XA + c\bar{c}}(b)}{dy_{c}dy_{\bar{c}}} 
\times \frac{d^2P_{pA \rightarrow XA +b\bar{b}}(b)}{dy_{b}dy_{\bar{b}}} \,. \label{DPS-1}
\end{eqnarray}
Here, $R_A$ and $R_p$ are the nuclei and the proton radii, respectively, $\vec b$ is the relative impact parameter 
($b\equiv |{\vec b}|$). The $\Theta$-function represents an approximate absorption factor that ensures that 
one considers only peripheral collisions when no nucleus break-up occurs \cite{PhysRevC.82.014904}. 
Let us consider the ingredients of the DPS cross section (\ref{DPS-1}) in detail.

%%%%%%%%%%%%%%%%%%%%%%%%%%%%
\subsection{SPS sub-process cross section}
\label{Sect:QQ-SPS}
%%%%%%%%%%%%%%%%%%%%%%%%%%%%

The differential probabilities $P(b)$ in Eq.~(\ref{DPS-1}) can be deduced from the corresponding SPS 
cross sections. For instance, for SPS production of $c\bar{c}$ pair we have
\begin{align}
\frac{d^3\sigma_{pA \rightarrow XA + c\bar{c}}}{dy_{c}dy_{\bar{c}}dp_\perp^2} 
& = \int d^2 \vec{b}\, \Theta(b - R_A - R_p) 
\frac{d^3 P_{pA \rightarrow XA + c\bar{c}}(b)}{dy_{c}dy_{\bar{c}}dp_\perp^2} 
\end{align}
where 
\begin{align}
\frac{d^3 P_{pA \rightarrow XA + c\bar{c}}(b)}{dy_{c}dy_{\bar{c}}dp_\perp^2} 
= \int d^2 \vec{b}_\gamma d^2 \vec{b}_g  \, \delta^{(2)} (\vec{b} + \vec{b}_g - \vec{b}_\gamma)
\int d \xi d x 
\,N_\gamma(\xi,\vec{b}_\gamma) G_g(x, \vec{b}_g) 
{\cal J}\,\frac{d^3\hat{\sigma}_{\gamma g \rightarrow c\bar{c}}}{dy_{c}dy_{\bar{c}}d\hat{t}} \,
\end{align}
in terms of the differential parton-level $\gamma+g \to c\bar{c}$ cross section, 
\begin{align}
&&\frac{d^3\hat{\sigma}_{\gamma g \rightarrow c\bar{c}}}{dy_{c}dy_{\bar{c}}d\hat{t}}
 = \frac{d\hat{\sigma}_{\gamma g \rightarrow Q\bar{Q}}}{d\hat{t}}
 \, \delta\left (y_{c} - \frac12 \ln \left( \frac{\xi}{x} \frac{\hat{u}}{\hat{t}} \right) \right) 
 \, \delta\left (y_{\bar{c}} - \frac12 \ln \left( \frac{\xi}{x} \frac{\hat{t}}{\hat{u}} \right) \right), 
 \end{align}
 written with the help of the modified Mandelstam variable
\begin{align}
\hat{t}= (p_c - p_\gamma)^2 - m^2_Q = - \sqrt{\hat{s}}\bigg(\frac{\sqrt{\hat{s}}}{2}-
\sqrt{\frac{\hat{s}}{4}-m^2_Q-p^2_\perp} \bigg)\,,
\end{align}
and the Jacobian is
\begin{align}
{\cal J}\equiv \Big|\frac{d\hat{t}}{dp^2_\perp}\Big|=
\frac{\sqrt{\hat{s}}}{2\sqrt{\frac{\hat{s}}{4}-m^2_Q-p^2_\perp}} \,,
\end{align}
as well as the photon $N_\gamma(\xi,\vec{b}_\gamma)$ and gluon $G_g(x,\vec{b}_g)$ distributions for 
the longitudinal momentum fractions $\xi$, $x$ and impact parameters $\vec{b}_\gamma$ 
and $\vec{b}_g$, respectively. The elementary cross section for the direct (fusion) sub-process reads 
in terms of the Mandelstam variables of the sub-process
\begin{eqnarray}
\frac{d^2\hat{\sigma}_{\gamma g \rightarrow Q\bar{Q}}}{d\hat{t}d\hat{u}} = \frac{\pi \alpha_s \alpha e^2_Q}{\hat{s}^2} \bigg[ \frac{\hat{t}}{\hat{u}}+
\frac{\hat{u}}{\hat{t}}+\frac{4m^2_Q\hat{s}}{\hat{t}\hat{u}}\bigg(1-\frac{m^2_Q\hat{s}}{\hat{t}\hat{u}}\bigg) \bigg] \delta(\hat{s}+\hat{t}+\hat{u}) \,.
\end{eqnarray}
Provided that the quark mass regulates the infrared behaviour of the integrals, there is no need to introduce additional low-$p_t$ cuts in order to unitarise the probabilities $P_{pA \rightarrow Q\bar{Q}}$ as for 
the heavy quarks they are below unity.

The standard WW photon flux is determined as
\begin{eqnarray}
\frac{d^3N_\gamma(\omega,{\vec b})}{d\omega d^2{\vec b}}= \frac{Z^2\alpha k^2}{\pi^2\omega b^2} \Big[ K^2_1(k)+\frac{1}{\gamma^2}K^2_0(k) \Big]\,, 
\qquad k=\frac{b\, \omega}{\gamma} \,,
\end{eqnarray}
with the nucleus charge $Z$, the modified Bessel functions of the second kind $K_0$ and $K_1$, the fine structure constant $\alpha$, the photon energy $\omega$, the Lorentz factor $\gamma$ defined as, 
$\gamma = \sqrt{s} / 2 m_p$, where $s$ is the center-of-mass (c.m.) energy per nucleon, and the proton mass $m_p=0.938$ GeV. 
For instance, at LHC $pA$ 2016 run (with $\sqrt{s} = 8.16$\,TeV) we have $\gamma_{\rm Pb} \approx 4350$, while for RHIC, 
$\gamma_{\rm Au} \approx 107$. For FCC collider (with $\sqrt{s} = 50$\,TeV), we have $\gamma \approx 26652$. Since we would like 
to work with the photon momentum fraction instead of photon energy, we then have
\begin{eqnarray}
\frac{d^3 N_\gamma(\xi,{\vec b})}{d\xi d^2{\vec b}}= 
\frac{\sqrt{s}}{2} \frac{d^3N_\gamma(\omega,{\vec b})}{d\omega d^2{\vec b}} 
\qquad \text{with} \qquad 
\xi = \frac{2 \omega}{\sqrt{s}} \,.
\end{eqnarray}

Another important ingredient is the impact-parameter dependent gluon distribution $G_g (x, \vec{b})$ 
that is often used in a factorised form,
\begin{align}
G_g(x, {\vec b}) = g(x)\, f_g({\vec b}) \,,
\label{Gg}
\end{align}
where $g(x)$ is the usual integrated gluon PDF, with an implicit factorisation scale dependence,
and $f_g (b)$ is the normalised spatial gluon distribution in the transverse plane
\begin{align}
f_g ({\vec b}) = \frac{\Lambda^2}{2 \pi} \frac{\Lambda  b}{2} K_1(\Lambda  b) \,, \qquad \int d^2 {\vec b} \, f_g ({\vec b}) = 1 \,.
\label{fg}
\end{align}
as in Ref.~\cite{Frankfurt:2010ea}. Here, $\Lambda \approx 1.5$ GeV and CT14nlo~\cite{Dulat:2015mca} collinear parton distributions are used with $\mu_F = \hat{s}$. 

\begin{figure}[!ht]
\centering
\includegraphics[width=.8\textwidth]{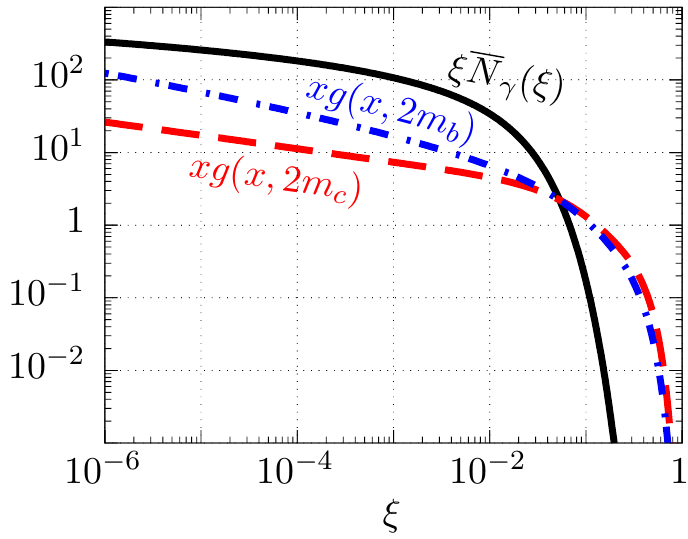}
\caption{The number distribution of photons $\overline{N}_\gamma(\xi)$ (times $\xi$) 
outside the nucleus that can interact with gluons in the proton in the 
considering photoproduction process. For comparison, the gluon distributions at two factorization scales $2m_c$ and $2 m_b$.}
\label{fig:Nbar}
\end{figure}

Consequently, the cross section for $c\bar c$ production in the SPS UPCs is related to the parton-level $\gamma+g \to c\bar{c}$ 
cross section as follows
\begin{eqnarray}
\nonumber
\frac{d^2\sigma_{pA \rightarrow XA + c\bar{c}}}{dy_{c}dy_{\bar{c}}} 
& = & \int d^2 \vec{b} d^2 \vec{b}_\gamma d^2 \vec{b}_g d \xi d x\, \Theta(b - R_A - R_p) 
\delta^{(2)} (\vec{b} + \vec{b}_g - \vec{b}_\gamma) \\
& \times & 
\,N_\gamma(\xi,\vec{b}_\gamma) G_g(x, \vec{b}_g) 
\frac{d^2\hat{\sigma}_{\gamma g \rightarrow c\bar{c}}}{dy_{c}dy_{\bar{c}}} \, .
\end{eqnarray}
This expression can be rewritten in the following equivalent form,
\begin{eqnarray}
\frac{d^2\sigma_{pA \rightarrow XA + c\bar{c}}}{dy_{c}dy_{\bar{c}}} 
= \int d \xi \int d x 
\, \overline{N}_\gamma(\xi) g(x)
\frac{d^2\hat{\sigma}_{\gamma g \rightarrow c\bar{c}}}{dy_{c}dy_{\bar{c}}} \,
 \int d^2 \vec{b}\, \Theta(b - R_A - R_p) T_{g\gamma} (\xi, \vec{b}) \,,
\label{SPS}
\end{eqnarray}
with the overlap function, that encapsulates all the impact parameter dependence in the matrix element squared, 
defined as follows
\begin{align}
\label{eq:Tg}
T_{g\gamma} (\xi, \vec{b})& = \frac{1}{\overline{N}_\gamma(\xi)} \int d^2 \vec{b}_\gamma  \,
\Theta(b_\gamma - R_A) N_\gamma(\xi,\vec{b}_\gamma) f_g(\vec{b} - \vec{b}_\gamma) \,,
\end{align} 
where
\begin{align}
\label{eq:Nbar}
\overline{N}_\gamma(\xi) = \int d^2 b \, \Theta(b - R_A) N_\gamma (\xi,\vec b) \,,
\end{align}
is the number distribution of photons that can interact in the considering process (outside the nucleus) shown in Fig.~{\ref{fig:Nbar}}. Here, it becomes apparent that the photon distribution is strongly peaked at low $\xi < 10^{-2}$. The distribution calculated with the WW flux is independent of energy if the Lorentz factor is very large, $\gamma \rightarrow \infty$.
%%%%%%%%%%%%%%%%%%%%%%%%%%%%%%%%%%
%%%%%%%%%%%%%%%%%%%%%%%%%%%%%%%%%%

It is worth noticing that the last integral in Eq.~(\ref{SPS}) can be transformed as follows
\begin{eqnarray*}
&& \int d^2 \vec{b} \Theta(b - R_A - R_p) T_{g\gamma} (\xi, \vec{b}) \\
&& \qquad \quad = \int d^2 \vec{b} \frac{\Theta(b - R_A - R_p)}{\overline{N}_\gamma(\xi)} \int d^2 \vec{b}_\gamma\,
\Theta(b_\gamma - R_A) N_\gamma(\xi,\vec{b}_\gamma) f_g(\vec{b}_\gamma - \vec{b}) \\ 
&& \qquad \quad =1 - \int d^2 \vec{b} \int d^2 \vec{b}_\gamma\,
\frac{\Theta(b_\gamma - R_A) ( 1 - \Theta(b - R_A - R_p))}{\overline{N}_\gamma(\xi)}   
N_\gamma(\xi,\vec{b}_\gamma) f_g(\vec{b}_\gamma - \vec{b}) \,.
\end{eqnarray*}
This relation explicitly demonstrates that, if the gluon distribution is very localized, i.e. $R_p \rightarrow 0$ 
together with $b_\gamma \rightarrow b$, the second term vanishes leaving no traces 
about the impact parameter dependence in the SPS cross section (\ref{SPS}).

For completeness, besides the direct production process whose formalism is discussed above, 
we have also included a sub-dominant resolved contribution following Ref.~\cite{Klein:2002wm}. 
By including the resolved component with the gluon-initiated hard subprocess $gg\to Q\bar Q$ at leading order (LO), as an example, 
one enables to pick a gluon (with momentum fraction $z$) from the incident photon by means of a gluon PDF 
in the photon, $g^\gamma(z,\mu^2)$, while the photon remnant hadronises into an unobserved hadronic system.
The corresponding contributions reads
\begin{eqnarray}
\nonumber
\frac{d^2\sigma^\text{Resolved, gluon}_{pA \rightarrow XA + c\bar{c}}}{dy_{c}dy_{\bar{c}}} 
& = & \int d \xi d x  \overline{N}_\gamma(\xi) g(x) \int d z  g^{\gamma}(z) \, 
\frac{d^2\hat{\sigma}_{gg \rightarrow Q\bar{Q}}}{dy_{c}dy_{\bar{c}}} \int d^2 \vec{b}\, 
\Theta(b - R_A - R_p) T_{g\gamma} (\xi, \vec{b}) \,.
\label{SPS-Resolved}
\end{eqnarray}
In our numerical analysis, we also include the second relevant subprocess $q \bar q\to Q\bar Q$ at LO. 
The PDFs in the photon are taken from Ref.~\cite{Aurenche:2005da}. Such a resolved photon 
contribution increases the differential cross section only slightly and mostly 
at negative rapidities (for more details and the corresponding figures, see below).

%%%%%%%%%%%%%%%%%%%%%%%%%%%%%%%%%%%%%%%%%%%%%%%%%%%%%%%%%%%%%%%%%%%
\subsection{Pocket formula for the DPS cross section in $pA$ UPCs}
\label{Sect:DPS-pocket}
%%%%%%%%%%%%%%%%%%%%%%%%%%%%%%%%%%%%%%%%%%%%%%%%%%%%%%%%%%%%%%%%%%%

Consider now the formalism for the DPS mechanism of direct $c\bar{c}b\bar{b}$ production in $pA$ UPCs 
(while the resolved photon contributions to the DPS are also included into the numerical analysis).

In what follows, we define the nucleus with positive rapidity, while the proton with negative one, 
such that the longitudinal photon and gluon momentum fractions are given by
\begin{eqnarray}
    \xi_i = \frac{m_{i,\perp}}{\sqrt{s}} ( e^{y_{Q_i}} + e^{y_{\bar Q_i} } ) \,, \qquad 
    x_i = \frac{m_{i,\perp}}{\sqrt{s}} ( e^{-y_{Q_i}} + e^{-y_{\bar Q_i} } ) \,, \qquad 
    m^2_{i,\perp} = m_{Q_i}^2 + p_{i,\perp}^2 \,,
\end{eqnarray}
respectively, where index $i=1,2$ denotes the elementary SPS processes or, equivalently, 
the heavy quark species the photon and gluon are coupled to, namely, $Q_{1,2}\equiv c,b$ 
in the considering process, $m_{i,\perp}$ is the transverse mass of the heavy quark, $y_{Q_i}$ ($y_{\bar Q_i}$) is 
the heavy quark (anti-quark) rapidity. Then the invariant mass of each $Q_i\bar Q_i$ pair reads 
\begin{align}
M^2_{i,Q\bar{Q}}= 2x_i\xi_i s = 2m^2_{i,\perp}(1+\cosh(y_{Q_i}-y_{\bar Q_i}))
\end{align}
such that
\begin{eqnarray}
dx_id\xi_i=\frac{m^2_{i,\perp}}{s}\big|\sinh(y_{Q_i}-y_{\bar Q_i})\big|dy_{Q_i}dy_{\bar Q_i} \,.
\end{eqnarray}

Consequently, the DPS cross section (\ref{DPS-1}) can be rewritten in terms of parton-level elementary 
photon-gluon fusion cross sections as
\begin{eqnarray}
\nonumber
&& \frac{d^4\sigma_{pA \rightarrow XA + c\bar{c} + b\bar{b}}}{dy_{c}dy_{\bar{c}}dy_{b}dy_{\bar{b}}}
= \int d^2 b \, \Theta(b - R_A - R_p) \int d^2 \vec{b}_{\gamma,1} \, \Theta (b_{\gamma,1} - R_A) 
\int d^2 \vec{b}_{\gamma,2} \, \Theta (b_{\gamma,2} - R_A) \\
&& \qquad\quad \times
\int d \xi_1 d \xi_2 d x_1 d x_2
N_{\gamma\gamma} (\xi_1, \vec{b}_{\gamma,1}; \xi_2, \vec{b}_{\gamma,2})  
G_{gg} (x_1, \vec{b}_{g,1}; x_2, \vec{b}_{g,2}) 
\frac{d^2\hat{\sigma}_{\gamma g \rightarrow c\bar{c}}}{dy_{c}dy_{\bar{c}}}
\frac{d^2\hat{\sigma}_{\gamma g \rightarrow b\bar{b}}}{dy_{b}dy_{\bar{b}}}
\end{eqnarray}
where $\vec{b}_{g,i} = \vec{b}_{\gamma,i} - \vec{b}$ for $ i = 1,2$, and $N_{\gamma\gamma}$ 
($G_{gg}$) is the corresponding di-photon (di-gluon) distribution. 

If we neglect any correlations between the individual photon and gluon exchanges, the di-photon 
and di-gluon distributions are conveniently represented in a factorised form, i.e.
\begin{eqnarray} \nonumber
    N_{\gamma\gamma} (\xi_1, \vec{b}_{\gamma,1}; \xi_2, \vec{b}_{\gamma,2}) & = & N_\gamma (\xi_1, \vec{b}_{\gamma,1}) 
    N_\gamma (\xi_2, \vec{b}_{\gamma,2}) \\
    G_{gg} (x_1, \vec{b}_{g,1}; x_2, \vec{b}_{g,2}) & = & G_g (x_1, \vec{b}_{g,1}) G_g (x_2, \vec{b}_{g,2}) \,,
    \label{factorisation}
\end{eqnarray}
in terms of the quasi-real single photon $N_\gamma (\xi, \vec{b})$ and gluon $G_g(x,\vec{b})$ distributions 
defined above. Note, the above factorisation formulas are approximations valid for $\xi_{1,2},x_{1,2}\ll 1$ only \cite{Blok:2011bu,Blok:2013bpa,Chang:2012nw,Gaunt:2009re,Rinaldi:2015cya}.

Using Eqs.~(\ref{factorisation}), (\ref{Gg}), (\ref{fg}) and (\ref{eq:Tg}), it is straightforward 
to transform the resulting DPS cross section to the following simple form
\begin{align} \label{DPS-main}
\frac{d^4\sigma_{pA \rightarrow XA + c\bar{c} + b\bar{b}}}{dy_{c}dy_{\bar{c}}dy_{b}dy_{\bar{b}}}
& = \int d \xi_1 d x_1 d \xi_2 d x_2 \,\, 
\frac{\overline{N}_\gamma(\xi_1) g(x_1)
\overline{N}_\gamma(\xi_2) g(x_2)}{\sigma_\text{eff}(\xi_1, \xi_2)}
\frac{d^2\hat{\sigma}_{\gamma g \rightarrow c\bar{c}}}{dy_{c}dy_{\bar{c}}} \,
\frac{d^2\hat{\sigma}_{\gamma g \rightarrow b\bar{b}}}{dy_{b}dy_{\bar{b}}} \,
\end{align}
where $\overline{N}_\gamma(\xi)$ is defined in Eq.~(\ref{eq:Nbar}),
and the definition of the effective cross section
\begin{align}
\sigma_\text{eff}(\xi_1, \xi_2) 
\equiv \Bigg[\int d^2 b \, \Theta(b - R_A - R_p)  T_{g\gamma}(\xi_1, b) T_{g\gamma}(\xi_2, b)\Bigg]^{-1} \,,
\label{Eq:eff}
\end{align}
has been introduced. This way we arrive at an analogue of the pocket formula applicable 
for the DPS contribution to $c\bar{c}b\bar{b}$ production in $pA$ UPCs. The equation 
(\ref{DPS-main}) is valid also for the DPS contribution to the $c\bar c c\bar c$ and 
$b\bar b b\bar b$ production processes apart from the change in the symmetry factor.

In what follows, we wish to investigate the corresponding SPS and DPS differential (in rapidity) 
cross sections making predictions for future measurements.

%%%%%%%%%%%%%%%%%%%%%%%%%%%%%
\section{Numerical results}
\label{Sect:Results}
%%%%%%%%%%%%%%%%%%%%%%%%%%%%%

In our numerical analysis of the SPS and DPS cross sections for heavy flavor production in 
$pA$ UPCs, we consider lead nucleus, with radius $R_{\rm Pb}=5.5$ fm (the proton radius 
is fixed to $R_p=0.87$ fm), while the heavy quark masses are taken to be $m_c=1.4$ GeV 
and $m_b=4.75$ GeV.
%%%%%%%%%%%%%%%%%%%%%%%%%%%%%%%%%%
\begin{figure}[!ht]
\centering
\includegraphics[width=.8\textwidth]{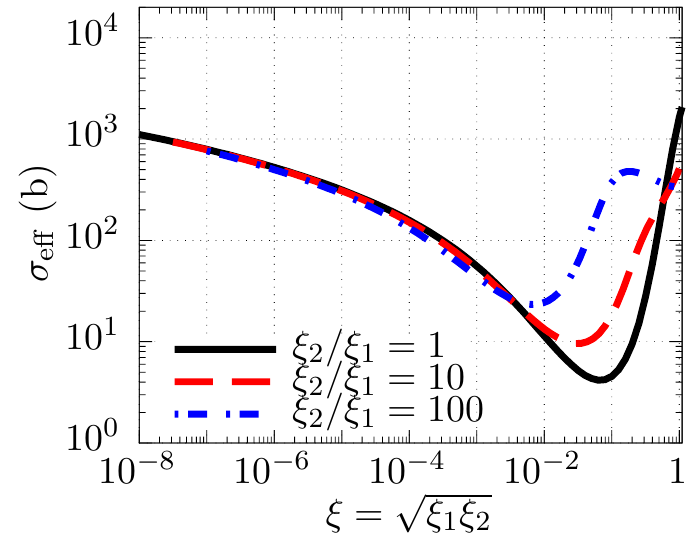}
\caption{The DPS effective cross section as a function of $\sqrt{\xi_1 \xi_2}$.}
\label{fig:sig-eff}
\end{figure}
%%%%%%%%%%%%%%%%%%%%%%%%%%%%%%%%%%

In Fig.~\ref{fig:sig-eff} the effective cross-section for pA UPCs of Eq.~\ref{Eq:eff} is plotted as a function of photon momentum fraction $\xi$. This plot carries essentially the information of where the photons are, but not of the number of photons outside the nucleus, as it is factored out in $\overline{N}_\gamma(\xi)$. The main contribution to this result arises when the two photons are inside the proton. With the model used here, the plot does not change with energy or factorization scale. 

Take first the case when $\xi_1 =\xi_2$. For small $\xi$, the photons are too spread and it is more rare that they overlap, then the effective cross section is larger. For large $\xi$, the photons are in a narrow shell just outside the nucleus, and if the width of this shell is smaller than the proton radius, it is clear that $\sigma_\text{eff}$ should also grow. That explains the minimum around $\xi \approx 0.07$, when half of the photons outside the nucleus have $b_\gamma - R_\text{Pb} < 1.0$\,fm, i.e., approximately the proton radius.

In the case of $\xi_2 / \xi_1 > 1$, $\sigma_\text{eff}$ can have two minima, as shown in the plot. That happens because the two photon distributions have their maximum probability of finding the photons inside the proton at different $\xi$.  
%%%%%%%%%%%%%%%%%%%%%%%%%%%%%%%%%%
\begin{figure*}[!htbt]
\begin{minipage}{0.495\textwidth}
 \centerline{\includegraphics[width=1.0\textwidth]{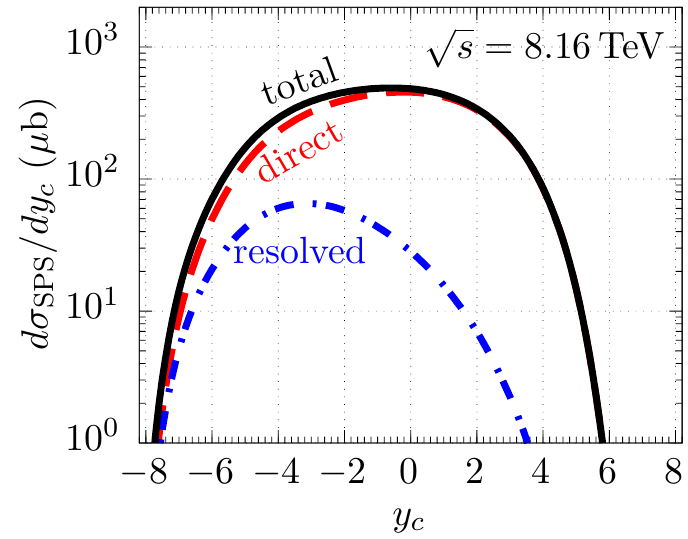}}
\end{minipage} \hfill
\begin{minipage}{0.495\textwidth}
 \centerline{\includegraphics[width=1.0\textwidth]{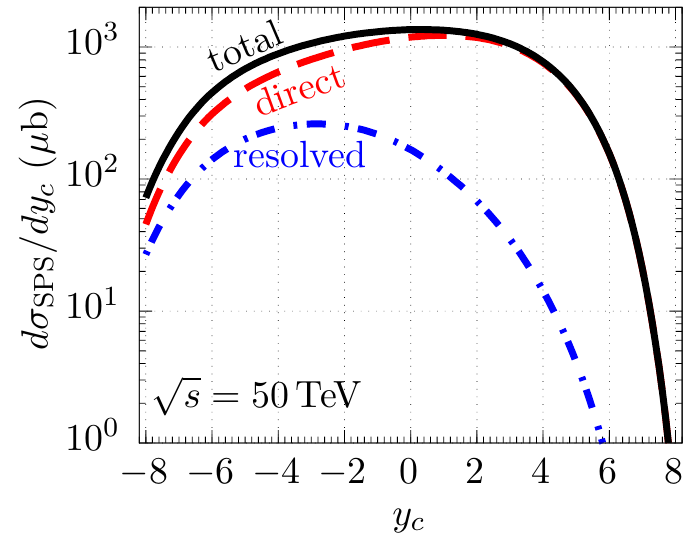}}
\end{minipage}
\caption{
The SPS $c\bar c$ quark production cross section in $pA$ UPCs as a function of rapidity $y_c$. Energy of $\sqrt{s}=8.16$\,TeV is shown in the left panel and $\sqrt{s}=50$\,TeV is shown in the right panel. Heavy ion is coming from the left. }
\label{fig:SPS}
\end{figure*}
%%%%%%%%%%%%%%%%%%%%%%%%%%%%%%%%%%

To better understand our double parton results, we recalculate the SPS cross section in Fig.~\ref{fig:SPS}. We show the differential cross section in $c$ quark rapidity, for energies of 8.16\,TeV and 50\,TeV. The heavy ion comes from the left, while the proton comes from the right. 

We detail the direct and the resolved contributions to the result. The resolved 
contribution is only relevant at small $z\xi$ and large $x$. 
Indeed, we see a harder decrease at positive than at negative rapidities, due to the nucleus photons 
having a sharper cutoff when $\xi \rightarrow 1$ than the proton gluons. The curves are almost 
flat at central rapidities, such that the resolved contribution makes it even flatter due to a small modification to the shape of the resulting differential cross section (mostly) in the negative rapidity domain. While the relative importance of the resolved contribution grows with energy, it remains to be minor compared to the direct one, see also Table~\ref{Tab:total}.
%%%%%%%%%%%%%%%%%%%%%%%%%%%%%%%%%%
\begin{figure*}[!ht]
\begin{minipage}{0.495\textwidth}
\centerline{\includegraphics[width=1.0\textwidth]{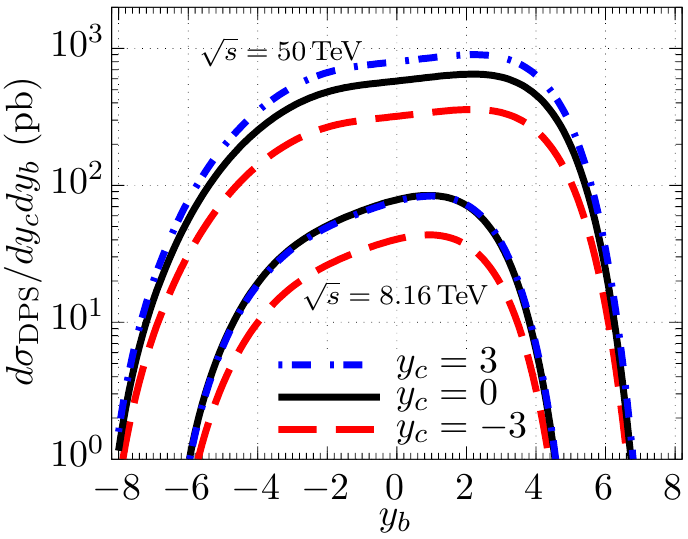}}
\end{minipage}\hfill 
\begin{minipage}{0.495\textwidth}
 \centerline{\includegraphics[width=1.0\textwidth]{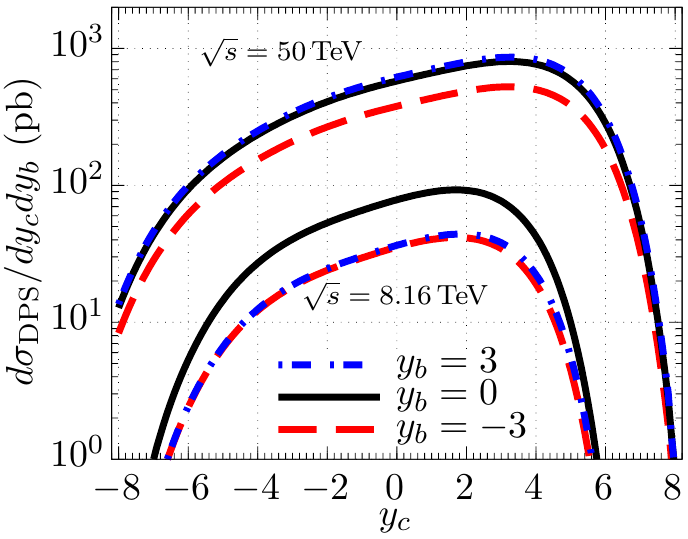}}
\end{minipage}
\caption{
The DPS $c\bar c b\bar b$ production cross section in $pA$ UPCs as a function of $c$-quark 
rapidity at fixed $b$ rapidity (left panel) and as a function of $b$-quark 
rapidity at fixed $c$-quark rapidity (right panel).}
\label{fig:DPS-ycb}
\end{figure*}
%%%%%%%%%%%%%%%%%%%%%%%%%%%%%%%%%%

For double parton scattering, we present, in Fig.~\ref{fig:DPS-ycb}, the $c\bar c b\bar b$ cross section with $y_{\bar c}, y_{\bar b}$ integrated out but differential in $y_c, y_b$ (at left and right panels, respectively). The production at central rapidities increases with rapidity; this is in contrast with the SPS case where it was flat. If the DPS was just the product of two SPS cross sections, we would not see such an increase. In effect, this is a result of the effective cross section in the denominator, that decreases as the photon energy fraction $\xi$ increases for $\xi < 0.07$. Therefore, this behaviour gives strength to our result that the pocket formula cannot have a constant effective cross section.
%%%%%%%%%%%%%%%%%%%%%%%%%%%%%%%%%%
\begin{figure*}[!ht]
\begin{minipage}{0.495\textwidth}
\centerline{\includegraphics[width=1.0\textwidth]{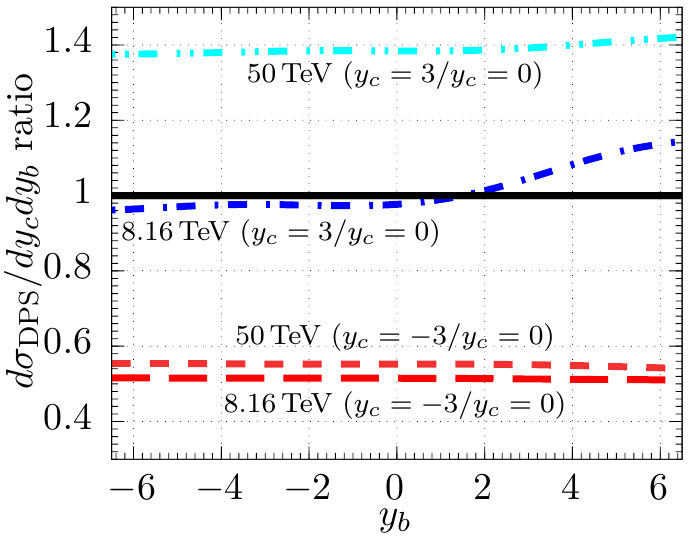}}
\end{minipage}\hfill 
\begin{minipage}{0.495\textwidth}
 \centerline{\includegraphics[width=1.0\textwidth]{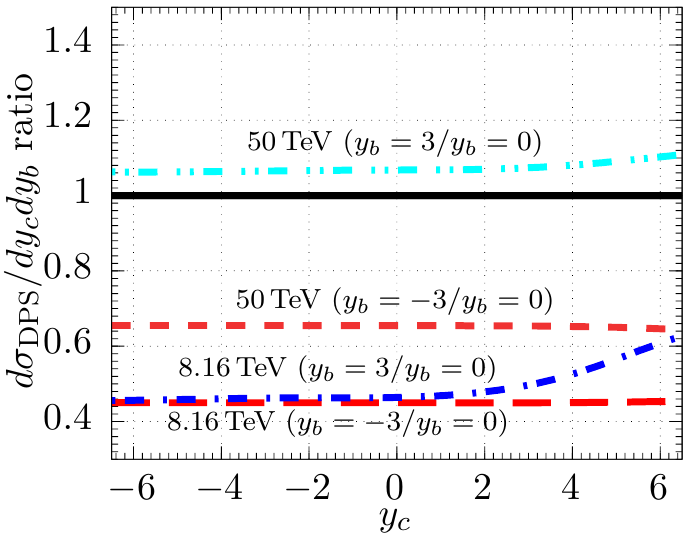}}
\end{minipage}
\caption{
The same as in Fig.~\ref{fig:DPS-ycb} but for the ratios of the differential cross section at each fixed rapidity to a reference curve at fixed $y_c=0$ (left panel) and $y_b=0$ (right panel) and at each given energy.}
\label{fig:ratio}
\end{figure*}
%%%%%%%%%%%%%%%%%%%%%%%%%%%%%%%%%%

As a matter of fact, any difference (other than a multiplicative factor) between Fig.~\ref{fig:DPS-ycb} and Fig.~\ref{fig:SPS} are due to the dependence of the effective cross section on $\xi$. In addition, the different lines in Fig.~\ref{fig:DPS-ycb} presenting different behaviours as the fixed rapidity is changed is a direct result of fact that the photon $b$ distribution changes with $\xi$. In order to illustrate how the shape changes with rapidity, we have added an extra Fig.~\ref{fig:ratio} showing the ratios of the differential cross section to a given reference curve at fixed $y_c=0$ (left panel) and $y_b=0$ (right panel) corresponding left and right panels of Fig.~\ref{fig:DPS-ycb}, respectively. As expected, the largest deviation in shapes emerge at forward rapidities corresponding to large $\xi$. It is much easier to see this effect in our UPC example than in standard four gluon DPS, since the photon impact parameter distributions have a clearer and stronger dependence on the longitudinal momentum fraction as opposed to gluons. Just to clarify, we remark again that no correlations between the two photons were taken into account, in the sense that picking the first photon does not changes the distribution of the second photon. 
%%%%%%%%%%%%%%%%%%%%%%%%%%%%%%%%%%
\begin{table}{}
	\begin{center}
		\begin{tabular}{|r|c|c|c|}
\hline
$\sqrt{s}$\,(TeV)   &  8.16 &  50  &  100\\
\hline
\multicolumn{4}{|c|}{SPS UPC $c \bar{c}$ production in mb}\\
\hline
Direct   & \, 3.10 \, & \, 10.46 \, & \, 15.75 \, \\
Resolved &  0.35  &  1.81 &  3.03 \\
Total 	 &  3.45  & 12.27 & 18.78 \\ 
 \hline
\multicolumn{4}{|c|}{DPS UPC $c \bar{c} b \bar{b}$ production in nb}\\
\hline
Total   &  3.55  & 54.1 & 136
 \\
\hline 
\end{tabular}
\vspace{0.04cm}
\caption{Table with the integrated cross sections for DPS and SPS production processes.}
\label{Tab:total}
\end{center} 
\end{table}
%%%%%%%%%%%%%%%%%%%%%%%%%%%%%%%%%%

%%%%%%%%%%%%%%%%%%%%%%%%%%%%%%%%%%
\begin{figure*}[!ht]
\begin{minipage}{0.495\textwidth}
 \centerline{\includegraphics[width=1.0\textwidth]{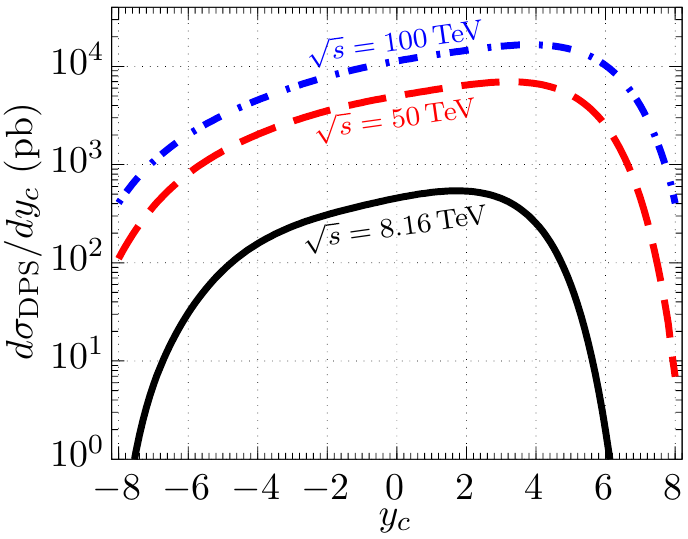}}
\end{minipage}\hfill
\begin{minipage}{0.495\textwidth}
 \centerline{\includegraphics[width=1.0\textwidth]{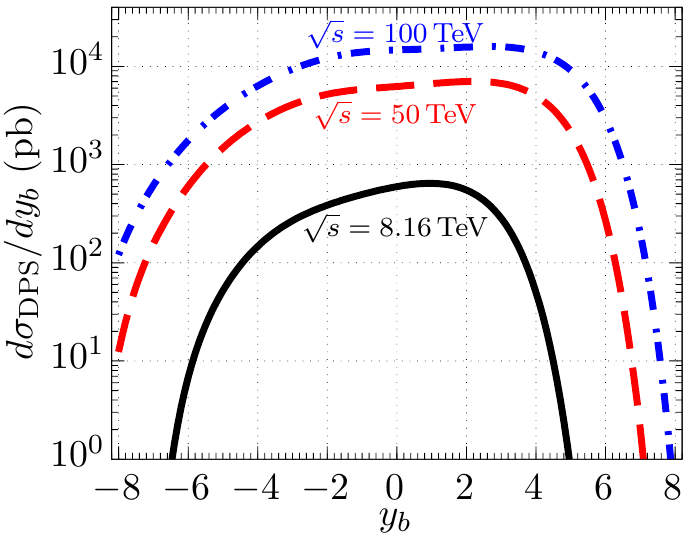}}
\end{minipage}
\caption{DPS $c\bar c b \bar b$ production cross section in $pA$ UPCs as a function of $c$ rapidity with $y_b$ integrated (left panel) and as a function of $b$-quark rapidity with $y_c$ integrated (right panel).}
\label{fig:DPS-int}
\end{figure*}
%%%%%%%%%%%%%%%%%%%%%%%%%%%%%%%%%%

In Fig.~\ref{fig:DPS-int} we integrate over one more rapidity, leaving only $y_b$ or $y_c$ unintegrated. Together with Table~\ref{Tab:total}, we see that we have a significant cross section of the order of nanobarns, indicating that such observable can be measured currently at the LHC and of course also at higher energy future colliders. 

In the differential cross sections, e.g.\ shown in Figs.~\ref{fig:SPS}, \ref{fig:DPS-ycb}, and \ref{fig:DPS-int}, integrated over the antiquark rapidities, one can increase statistics by multiplying them by a factor of two (for SPS) or by the factor of four (for DPS) if a measurement detects open heavy flavored mesons containing both heavy quarks and antiquarks. % and is capable of distinguishing them.
In this case, charged $D^\pm$ and $B^\pm$ mesons should be detected in the DPS final state, simultaneously ensuring that $D^+D^-$ and $B^+B^-$ meson pairs are produced at well separated rapidity domains to suppress the SPS $\gamma+g \to c\bar c b \bar b$ background contribution. In principle, for this purpose it suffices to consider the individual (anti)quark rapidities $y_c$ (or $y_{\bar c}$) and $y_b$ (or $y_{\bar b}$) to be far apart from each other.

%%%%%%%%%%%%%%%
\section{Conclusions}
\label{Sect:Concl}
%%%%%%%%%%%%%%%

In this paper we investigated the double parton interaction between heavy-ion and proton 
in ultraperipheral collisions. For the main contribution, where two photons from the heavy ion 
interact with two gluons from the proton, we developed a new pocket formula, with some peculiarities when compared to the usual one. In our case, as the distribution of photons is not as localized in impact parameter as the gluon distribution, the effective cross section is rather large, roughly dozens of barns. Another consequence is that the effective cross section is heavily dependent on the photon longitudinal momentum fraction, and this cannot be neglected. Therefore, we do not have a simple multiplication of SPS cross sections, but instead a convolution in the photon longitudinal momentum fraction. 

We presented our results in terms of the cross section to produce $c$ and $b$ quarks as a function of rapidities. In this way, we can assert that each heavy quark gives us information about one of the gluons in the initial state. Therefore, this is an effective and direct probe of the double gluon distribution that can be studied at the HL-LHC or at a future collider, e.g. at 50 TeV, for which the predictions are shown in Figs.~\ref{fig:DPS-ycb} and \ref{fig:DPS-int}.

We point out that the most efficient way of suppressing the SPS $\gamma+g \to c\bar c b \bar b$ 
background contribution is to measure the open charm and open bottom mesons at large relative rapidity separation of a few units. So, future measurements aiming at precision measurement of DPS contribution 
in the considered process are encouraged to have the corresponding detectors covering different rapidity
(for example, in central and forward/backward) domains of the phase space.

%--------------------------
\section*{Acknowledgments}
%--------------------------

This work was supported by Fapesc, INCT-FNA (464898/2014-5), and CNPq (Brazil) for EGdO and EH. This study was financed in part by the Coordena\c{c}\~ao 
de Aperfei\c{c}oamento de Pessoal de N\'ivel Superior -- Brasil (CAPES) -- Finance Code 001. The work has been performed in the framework of COST Action CA15213 
``Theory of hot matter and relativistic heavy-ion collisions'' (THOR). R.P.~is supported in part by the Swedish Research Council grants, contract numbers 621-2013-4287 
and 2016-05996, by CONICYT grant MEC80170112, as well as by the European Research Council (ERC) under the European Union's Horizon 2020 research and innovation 
programme (grant agreement No 668679). This work was also supported in part by the Ministry of Education, Youth and Sports of the Czech Republic, project LT17018.

%------------------------
\bibliography{bib-DPS}

\begin{thebibliography}{49}
\expandafter\ifx\csname natexlab\endcsname\relax\def\natexlab#1{#1}\fi
\expandafter\ifx\csname bibnamefont\endcsname\relax
  \def\bibnamefont#1{#1}\fi
\expandafter\ifx\csname bibfnamefont\endcsname\relax
  \def\bibfnamefont#1{#1}\fi
\expandafter\ifx\csname citenamefont\endcsname\relax
  \def\citenamefont#1{#1}\fi
\expandafter\ifx\csname url\endcsname\relax
  \def\url#1{\texttt{#1}}\fi
\expandafter\ifx\csname urlprefix\endcsname\relax\def\urlprefix{URL }\fi
\providecommand{\bibinfo}[2]{#2}
\providecommand{\eprint}[2][]{\url{#2}}

\bibitem[{\citenamefont{Paver and Treleani}(1982)}]{Paver1982}
\bibinfo{author}{\bibfnamefont{N.}~\bibnamefont{Paver}} \bibnamefont{and}
  \bibinfo{author}{\bibfnamefont{D.}~\bibnamefont{Treleani}},
  \bibinfo{journal}{Il Nuovo Cimento A (1965-1970)}
  \textbf{\bibinfo{volume}{70}}, \bibinfo{pages}{215} (\bibinfo{year}{1982}),
  ISSN \bibinfo{issn}{1826-9869},
  \urlprefix\url{https://doi.org/10.1007/BF02814035}.

\bibitem[{\citenamefont{Mekhfi}(1985)}]{Mekhfi:1983az}
\bibinfo{author}{\bibfnamefont{M.}~\bibnamefont{Mekhfi}},
  \bibinfo{journal}{Phys. Rev.} \textbf{\bibinfo{volume}{D32}},
  \bibinfo{pages}{2371} (\bibinfo{year}{1985}).

\bibitem[{\citenamefont{Sj\"ostrand and van Zijl}(1987)}]{PhysRevD.36.2019}
\bibinfo{author}{\bibfnamefont{T.}~\bibnamefont{Sj\"ostrand}} \bibnamefont{and}
  \bibinfo{author}{\bibfnamefont{M.}~\bibnamefont{van Zijl}},
  \bibinfo{journal}{Phys. Rev. D} \textbf{\bibinfo{volume}{36}},
  \bibinfo{pages}{2019} (\bibinfo{year}{1987}),
  \urlprefix\url{https://link.aps.org/doi/10.1103/PhysRevD.36.2019}.

\bibitem[{\citenamefont{Gaunt and Stirling}(2010)}]{Gaunt:2009re}
\bibinfo{author}{\bibfnamefont{J.~R.} \bibnamefont{Gaunt}} \bibnamefont{and}
  \bibinfo{author}{\bibfnamefont{W.~J.} \bibnamefont{Stirling}},
  \bibinfo{journal}{JHEP} \textbf{\bibinfo{volume}{03}}, \bibinfo{pages}{005}
  (\bibinfo{year}{2010}), \eprint{0910.4347}.

\bibitem[{\citenamefont{Diehl et~al.}(2012)\citenamefont{Diehl, Ostermeier, and
  Schafer}}]{Diehl:2011yj}
\bibinfo{author}{\bibfnamefont{M.}~\bibnamefont{Diehl}},
  \bibinfo{author}{\bibfnamefont{D.}~\bibnamefont{Ostermeier}},
  \bibnamefont{and} \bibinfo{author}{\bibfnamefont{A.}~\bibnamefont{Schafer}},
  \bibinfo{journal}{JHEP} \textbf{\bibinfo{volume}{03}}, \bibinfo{pages}{089}
  (\bibinfo{year}{2012}), \bibinfo{note}{[Erratum: JHEP03,001(2016)]},
  \eprint{1111.0910}.

\bibitem[{\citenamefont{Manohar and Waalewijn}(2012)}]{PhysRevD.85.114009}
\bibinfo{author}{\bibfnamefont{A.~V.} \bibnamefont{Manohar}} \bibnamefont{and}
  \bibinfo{author}{\bibfnamefont{W.~J.} \bibnamefont{Waalewijn}},
  \bibinfo{journal}{Phys. Rev. D} \textbf{\bibinfo{volume}{85}},
  \bibinfo{pages}{114009} (\bibinfo{year}{2012}),
  \urlprefix\url{https://link.aps.org/doi/10.1103/PhysRevD.85.114009}.

\bibitem[{\citenamefont{Aaij et~al.}(2016)}]{Aaij:2015wpa}
\bibinfo{author}{\bibfnamefont{R.}~\bibnamefont{Aaij}} \bibnamefont{et~al.}
  (\bibinfo{collaboration}{LHCb}), \bibinfo{journal}{JHEP}
  \textbf{\bibinfo{volume}{07}}, \bibinfo{pages}{052} (\bibinfo{year}{2016}),
  \eprint{1510.05949}.

\bibitem[{\citenamefont{Blok and Strikman}(2016)}]{Blok:2016lmd}
\bibinfo{author}{\bibfnamefont{B.}~\bibnamefont{Blok}} \bibnamefont{and}
  \bibinfo{author}{\bibfnamefont{M.}~\bibnamefont{Strikman}},
  \bibinfo{journal}{Eur. Phys. J.} \textbf{\bibinfo{volume}{C76}},
  \bibinfo{pages}{694} (\bibinfo{year}{2016}), \eprint{1608.00014}.

\bibitem[{\citenamefont{Rinaldi et~al.}(2016)\citenamefont{Rinaldi, Scopetta,
  Traini, and Vento}}]{Rinaldi:2015cya}
\bibinfo{author}{\bibfnamefont{M.}~\bibnamefont{Rinaldi}},
  \bibinfo{author}{\bibfnamefont{S.}~\bibnamefont{Scopetta}},
  \bibinfo{author}{\bibfnamefont{M.}~\bibnamefont{Traini}}, \bibnamefont{and}
  \bibinfo{author}{\bibfnamefont{V.}~\bibnamefont{Vento}},
  \bibinfo{journal}{Phys. Lett.} \textbf{\bibinfo{volume}{B752}},
  \bibinfo{pages}{40} (\bibinfo{year}{2016}), \eprint{1506.05742}.

\bibitem[{\citenamefont{Maciu\l{}a and Szczurek}(2018)}]{PhysRevD.97.094010}
\bibinfo{author}{\bibfnamefont{R.}~\bibnamefont{Maciu\l{}a}} \bibnamefont{and}
  \bibinfo{author}{\bibfnamefont{A.}~\bibnamefont{Szczurek}},
  \bibinfo{journal}{Phys. Rev. D} \textbf{\bibinfo{volume}{97}},
  \bibinfo{pages}{094010} (\bibinfo{year}{2018}),
  \urlprefix\url{https://link.aps.org/doi/10.1103/PhysRevD.97.094010}.

\bibitem[{\citenamefont{Abe et~al.}(1993)\citenamefont{Abe, Albrow, Amidei,
  Anway-Wiese, Apollinari, Atac, Auchincloss, Azzi, Baden, Bacchetta
  et~al.}}]{PhysRevD.47.4857}
\bibinfo{author}{\bibfnamefont{F.}~\bibnamefont{Abe}},
  \bibinfo{author}{\bibfnamefont{M.}~\bibnamefont{Albrow}},
  \bibinfo{author}{\bibfnamefont{D.}~\bibnamefont{Amidei}},
  \bibinfo{author}{\bibfnamefont{C.}~\bibnamefont{Anway-Wiese}},
  \bibinfo{author}{\bibfnamefont{G.}~\bibnamefont{Apollinari}},
  \bibinfo{author}{\bibfnamefont{M.}~\bibnamefont{Atac}},
  \bibinfo{author}{\bibfnamefont{P.}~\bibnamefont{Auchincloss}},
  \bibinfo{author}{\bibfnamefont{P.}~\bibnamefont{Azzi}},
  \bibinfo{author}{\bibfnamefont{A.~R.} \bibnamefont{Baden}},
  \bibinfo{author}{\bibfnamefont{N.}~\bibnamefont{Bacchetta}},
  \bibnamefont{et~al.}, \bibinfo{journal}{Phys. Rev. D}
  \textbf{\bibinfo{volume}{47}}, \bibinfo{pages}{4857} (\bibinfo{year}{1993}),
  \urlprefix\url{https://link.aps.org/doi/10.1103/PhysRevD.47.4857}.

\bibitem[{\citenamefont{Aaboud et~al.}(2019)}]{Aaboud:2018tiq}
\bibinfo{author}{\bibfnamefont{M.}~\bibnamefont{Aaboud}} \bibnamefont{et~al.}
  (\bibinfo{collaboration}{ATLAS}), \bibinfo{journal}{Phys. Lett.}
  \textbf{\bibinfo{volume}{B790}}, \bibinfo{pages}{595} (\bibinfo{year}{2019}),
  \bibinfo{note}{[Phys. Lett.790,595(2019)]}, \eprint{1811.11094}.

\bibitem[{\citenamefont{Calucci and Treleani}(1999)}]{PhysRevD.60.054023}
\bibinfo{author}{\bibfnamefont{G.}~\bibnamefont{Calucci}} \bibnamefont{and}
  \bibinfo{author}{\bibfnamefont{D.}~\bibnamefont{Treleani}},
  \bibinfo{journal}{Phys. Rev. D} \textbf{\bibinfo{volume}{60}},
  \bibinfo{pages}{054023} (\bibinfo{year}{1999}),
  \urlprefix\url{https://link.aps.org/doi/10.1103/PhysRevD.60.054023}.

\bibitem[{\citenamefont{Diehl and Gaunt}(2018)}]{Diehl:2017wew}
\bibinfo{author}{\bibfnamefont{M.}~\bibnamefont{Diehl}} \bibnamefont{and}
  \bibinfo{author}{\bibfnamefont{J.~R.} \bibnamefont{Gaunt}},
  \bibinfo{journal}{Adv. Ser. Direct. High Energy Phys.}
  \textbf{\bibinfo{volume}{29}}, \bibinfo{pages}{7} (\bibinfo{year}{2018}),
  \eprint{1710.04408}.

\bibitem[{\citenamefont{Chang et~al.}(2013{\natexlab{a}})\citenamefont{Chang,
  Manohar, and Waalewijn}}]{PhysRevD.87.034009}
\bibinfo{author}{\bibfnamefont{H.-M.} \bibnamefont{Chang}},
  \bibinfo{author}{\bibfnamefont{A.~V.} \bibnamefont{Manohar}},
  \bibnamefont{and} \bibinfo{author}{\bibfnamefont{W.~J.}
  \bibnamefont{Waalewijn}}, \bibinfo{journal}{Phys. Rev. D}
  \textbf{\bibinfo{volume}{87}}, \bibinfo{pages}{034009}
  (\bibinfo{year}{2013}{\natexlab{a}}),
  \urlprefix\url{https://link.aps.org/doi/10.1103/PhysRevD.87.034009}.

\bibitem[{\citenamefont{Rinaldi et~al.}(2013)\citenamefont{Rinaldi, Scopetta,
  and Vento}}]{PhysRevD.87.114021}
\bibinfo{author}{\bibfnamefont{M.}~\bibnamefont{Rinaldi}},
  \bibinfo{author}{\bibfnamefont{S.}~\bibnamefont{Scopetta}}, \bibnamefont{and}
  \bibinfo{author}{\bibfnamefont{V.}~\bibnamefont{Vento}},
  \bibinfo{journal}{Phys. Rev. D} \textbf{\bibinfo{volume}{87}},
  \bibinfo{pages}{114021} (\bibinfo{year}{2013}),
  \urlprefix\url{https://link.aps.org/doi/10.1103/PhysRevD.87.114021}.

\bibitem[{\citenamefont{Rinaldi et~al.}(2014)\citenamefont{Rinaldi, Scopetta,
  Traini, and Vento}}]{Rinaldi:2014ddl}
\bibinfo{author}{\bibfnamefont{M.}~\bibnamefont{Rinaldi}},
  \bibinfo{author}{\bibfnamefont{S.}~\bibnamefont{Scopetta}},
  \bibinfo{author}{\bibfnamefont{M.}~\bibnamefont{Traini}}, \bibnamefont{and}
  \bibinfo{author}{\bibfnamefont{V.}~\bibnamefont{Vento}},
  \bibinfo{journal}{JHEP} \textbf{\bibinfo{volume}{12}}, \bibinfo{pages}{028}
  (\bibinfo{year}{2014}), \eprint{1409.1500}.

\bibitem[{\citenamefont{Åkesson et~al.}(1987)\citenamefont{Åkesson, Albrow,
  Almehed, Benary, Bøggild, Botner, Breuker, Carter, Carter, Choi
  et~al.}}]{Akesson1987163}
\bibinfo{author}{\bibfnamefont{T.}~\bibnamefont{Åkesson}},
  \bibinfo{author}{\bibfnamefont{M.}~\bibnamefont{Albrow}},
  \bibinfo{author}{\bibfnamefont{S.}~\bibnamefont{Almehed}},
  \bibinfo{author}{\bibfnamefont{O.}~\bibnamefont{Benary}},
  \bibinfo{author}{\bibfnamefont{H.}~\bibnamefont{Bøggild}},
  \bibinfo{author}{\bibfnamefont{O.}~\bibnamefont{Botner}},
  \bibinfo{author}{\bibfnamefont{H.}~\bibnamefont{Breuker}},
  \bibinfo{author}{\bibfnamefont{A.}~\bibnamefont{Carter}},
  \bibinfo{author}{\bibfnamefont{J.}~\bibnamefont{Carter}},
  \bibinfo{author}{\bibfnamefont{Y.}~\bibnamefont{Choi}}, \bibnamefont{et~al.},
  \bibinfo{journal}{Zeitschrift für Physik C Particles and Fields}
  \textbf{\bibinfo{volume}{34}}, \bibinfo{pages}{163} (\bibinfo{year}{1987}),
  \bibinfo{note}{cited By 140},
  \urlprefix\url{https://www.scopus.com/inward/record.uri?eid=2-s2.0-25044475120&doi=10.1007%2fBF01566757&partnerID=40&md5=eed609a5f61d17f88a971070778a3381}.

\bibitem[{\citenamefont{Alitti et~al.}(1991)\citenamefont{Alitti, Ambrosini,
  Ansari, Autiero, Bareyre, Bertram, Blaylock, Bonamy, Borer, Bourliaud
  et~al.}}]{ALITTI1991145}
\bibinfo{author}{\bibfnamefont{J.}~\bibnamefont{Alitti}},
  \bibinfo{author}{\bibfnamefont{G.}~\bibnamefont{Ambrosini}},
  \bibinfo{author}{\bibfnamefont{R.}~\bibnamefont{Ansari}},
  \bibinfo{author}{\bibfnamefont{D.}~\bibnamefont{Autiero}},
  \bibinfo{author}{\bibfnamefont{P.}~\bibnamefont{Bareyre}},
  \bibinfo{author}{\bibfnamefont{I.}~\bibnamefont{Bertram}},
  \bibinfo{author}{\bibfnamefont{G.}~\bibnamefont{Blaylock}},
  \bibinfo{author}{\bibfnamefont{P.}~\bibnamefont{Bonamy}},
  \bibinfo{author}{\bibfnamefont{K.}~\bibnamefont{Borer}},
  \bibinfo{author}{\bibfnamefont{M.}~\bibnamefont{Bourliaud}},
  \bibnamefont{et~al.}, \bibinfo{journal}{Physics Letters B}
  \textbf{\bibinfo{volume}{268}}, \bibinfo{pages}{145 } (\bibinfo{year}{1991}),
  ISSN \bibinfo{issn}{0370-2693},
  \urlprefix\url{http://www.sciencedirect.com/science/article/pii/037026939190937L}.

\bibitem[{\citenamefont{{The LHCb collaboration}
  et~al.}(2012)\citenamefont{{The LHCb collaboration}, Aaij, Abellan~Beteta,
  Adeva, Adinolfi, Adrover, Affolder, Ajaltouni, Albrecht, Alessio
  et~al.}}]{Aaij2012}
\bibinfo{author}{\bibnamefont{{The LHCb collaboration}}},
  \bibinfo{author}{\bibfnamefont{R.}~\bibnamefont{Aaij}},
  \bibinfo{author}{\bibfnamefont{C.}~\bibnamefont{Abellan~Beteta}},
  \bibinfo{author}{\bibfnamefont{B.}~\bibnamefont{Adeva}},
  \bibinfo{author}{\bibfnamefont{M.}~\bibnamefont{Adinolfi}},
  \bibinfo{author}{\bibfnamefont{C.}~\bibnamefont{Adrover}},
  \bibinfo{author}{\bibfnamefont{A.}~\bibnamefont{Affolder}},
  \bibinfo{author}{\bibfnamefont{Z.}~\bibnamefont{Ajaltouni}},
  \bibinfo{author}{\bibfnamefont{J.}~\bibnamefont{Albrecht}},
  \bibinfo{author}{\bibfnamefont{F.}~\bibnamefont{Alessio}},
  \bibnamefont{et~al.}, \bibinfo{journal}{Journal of High Energy Physics}
  \textbf{\bibinfo{volume}{2012}}, \bibinfo{pages}{141} (\bibinfo{year}{2012}),
  ISSN \bibinfo{issn}{1029-8479},
  \urlprefix\url{https://doi.org/10.1007/JHEP06(2012)141}.

\bibitem[{\citenamefont{Abazov et~al.}(2010)\citenamefont{Abazov, Abbott,
  Abolins, Acharya, Adams, Adams, Aguilo, Alexeev, Alkhazov, Alton
  et~al.}}]{PhysRevD.81.052012}
\bibinfo{author}{\bibfnamefont{V.~M.} \bibnamefont{Abazov}},
  \bibinfo{author}{\bibfnamefont{B.}~\bibnamefont{Abbott}},
  \bibinfo{author}{\bibfnamefont{M.}~\bibnamefont{Abolins}},
  \bibinfo{author}{\bibfnamefont{B.~S.} \bibnamefont{Acharya}},
  \bibinfo{author}{\bibfnamefont{M.}~\bibnamefont{Adams}},
  \bibinfo{author}{\bibfnamefont{T.}~\bibnamefont{Adams}},
  \bibinfo{author}{\bibfnamefont{E.}~\bibnamefont{Aguilo}},
  \bibinfo{author}{\bibfnamefont{G.~D.} \bibnamefont{Alexeev}},
  \bibinfo{author}{\bibfnamefont{G.}~\bibnamefont{Alkhazov}},
  \bibinfo{author}{\bibfnamefont{A.}~\bibnamefont{Alton}}, \bibnamefont{et~al.}
  (\bibinfo{collaboration}{The D0 Collaboration}), \bibinfo{journal}{Phys. Rev.
  D} \textbf{\bibinfo{volume}{81}}, \bibinfo{pages}{052012}
  (\bibinfo{year}{2010}),
  \urlprefix\url{https://link.aps.org/doi/10.1103/PhysRevD.81.052012}.

\bibitem[{\citenamefont{Aad et~al.}(2015)}]{Aad:2014kba}
\bibinfo{author}{\bibfnamefont{G.}~\bibnamefont{Aad}} \bibnamefont{et~al.}
  (\bibinfo{collaboration}{ATLAS}), \bibinfo{journal}{Eur. Phys. J.}
  \textbf{\bibinfo{volume}{C75}}, \bibinfo{pages}{229} (\bibinfo{year}{2015}),
  \eprint{1412.6428}.

\bibitem[{\citenamefont{Aaij et~al.}(2017)}]{Aaij:2016bqq}
\bibinfo{author}{\bibfnamefont{R.}~\bibnamefont{Aaij}} \bibnamefont{et~al.}
  (\bibinfo{collaboration}{LHCb}), \bibinfo{journal}{JHEP}
  \textbf{\bibinfo{volume}{06}}, \bibinfo{pages}{047} (\bibinfo{year}{2017}),
  \bibinfo{note}{[Erratum: JHEP10,068(2017)]}, \eprint{1612.07451}.

\bibitem[{\citenamefont{Sirunyan et~al.}(2018)}]{Sirunyan:2017hlu}
\bibinfo{author}{\bibfnamefont{A.~M.} \bibnamefont{Sirunyan}}
  \bibnamefont{et~al.} (\bibinfo{collaboration}{CMS}), \bibinfo{journal}{JHEP}
  \textbf{\bibinfo{volume}{02}}, \bibinfo{pages}{032} (\bibinfo{year}{2018}),
  \eprint{1712.02280}.

\bibitem[{\citenamefont{Diehl and Schäfer}(2011)}]{DIEHL2011389}
\bibinfo{author}{\bibfnamefont{M.}~\bibnamefont{Diehl}} \bibnamefont{and}
  \bibinfo{author}{\bibfnamefont{A.}~\bibnamefont{Schäfer}},
  \bibinfo{journal}{Physics Letters B} \textbf{\bibinfo{volume}{698}},
  \bibinfo{pages}{389 } (\bibinfo{year}{2011}), ISSN \bibinfo{issn}{0370-2693},
  \urlprefix\url{http://www.sciencedirect.com/science/article/pii/S0370269311002863}.

\bibitem[{Bar(2011)}]{Bartalini:2011jp}
\emph{\bibinfo{title}{{Multi-Parton Interactions at the LHC}}}
  (\bibinfo{year}{2011}), \eprint{1111.0469}.

\bibitem[{\citenamefont{Bansal et~al.}(2014)}]{Bansal:2014paa}
\bibinfo{author}{\bibfnamefont{S.}~\bibnamefont{Bansal}} \bibnamefont{et~al.},
  in \emph{\bibinfo{booktitle}{{Workshop on Multi-Parton Interactions at the
  LHC (MPI @ LHC 2013) Antwerp, Belgium, December 2-6, 2013}}}
  (\bibinfo{year}{2014}), \eprint{1410.6664}.

\bibitem[{\citenamefont{\L{}uszczak et~al.}(2012)\citenamefont{\L{}uszczak,
  Maciu\l{}a, and Szczurek}}]{PhysRevD.85.094034}
\bibinfo{author}{\bibfnamefont{M.}~\bibnamefont{\L{}uszczak}},
  \bibinfo{author}{\bibfnamefont{R.}~\bibnamefont{Maciu\l{}a}},
  \bibnamefont{and} \bibinfo{author}{\bibfnamefont{A.}~\bibnamefont{Szczurek}},
  \bibinfo{journal}{Phys. Rev. D} \textbf{\bibinfo{volume}{85}},
  \bibinfo{pages}{094034} (\bibinfo{year}{2012}),
  \urlprefix\url{https://link.aps.org/doi/10.1103/PhysRevD.85.094034}.

\bibitem[{\citenamefont{{The LHCb collaboration}
  et~al.}(2014)\citenamefont{{The LHCb collaboration}, Aaij, Beteta, Adeva,
  Adinolfi, Adrover, Affolder, Ajaltouni, Albrecht, Alessio et~al.}}]{Aaij2014}
\bibinfo{author}{\bibnamefont{{The LHCb collaboration}}},
  \bibinfo{author}{\bibfnamefont{R.}~\bibnamefont{Aaij}},
  \bibinfo{author}{\bibfnamefont{C.~A.} \bibnamefont{Beteta}},
  \bibinfo{author}{\bibfnamefont{B.}~\bibnamefont{Adeva}},
  \bibinfo{author}{\bibfnamefont{M.}~\bibnamefont{Adinolfi}},
  \bibinfo{author}{\bibfnamefont{C.}~\bibnamefont{Adrover}},
  \bibinfo{author}{\bibfnamefont{A.}~\bibnamefont{Affolder}},
  \bibinfo{author}{\bibfnamefont{Z.}~\bibnamefont{Ajaltouni}},
  \bibinfo{author}{\bibfnamefont{J.}~\bibnamefont{Albrecht}},
  \bibinfo{author}{\bibfnamefont{F.}~\bibnamefont{Alessio}},
  \bibnamefont{et~al.}, \bibinfo{journal}{Journal of High Energy Physics}
  \textbf{\bibinfo{volume}{2014}}, \bibinfo{pages}{108} (\bibinfo{year}{2014}),
  ISSN \bibinfo{issn}{1029-8479},
  \urlprefix\url{https://doi.org/10.1007/JHEP03(2014)108}.

\bibitem[{\citenamefont{Maciu\l{}a and Szczurek}(2013)}]{PhysRevD.87.074039}
\bibinfo{author}{\bibfnamefont{R.}~\bibnamefont{Maciu\l{}a}} \bibnamefont{and}
  \bibinfo{author}{\bibfnamefont{A.}~\bibnamefont{Szczurek}},
  \bibinfo{journal}{Phys. Rev. D} \textbf{\bibinfo{volume}{87}},
  \bibinfo{pages}{074039} (\bibinfo{year}{2013}),
  \urlprefix\url{https://link.aps.org/doi/10.1103/PhysRevD.87.074039}.

\bibitem[{\citenamefont{Del~Fabbro and Treleani}(2002)}]{PhysRevD.66.074012}
\bibinfo{author}{\bibfnamefont{A.}~\bibnamefont{Del~Fabbro}} \bibnamefont{and}
  \bibinfo{author}{\bibfnamefont{D.}~\bibnamefont{Treleani}},
  \bibinfo{journal}{Phys. Rev. D} \textbf{\bibinfo{volume}{66}},
  \bibinfo{pages}{074012} (\bibinfo{year}{2002}),
  \urlprefix\url{https://link.aps.org/doi/10.1103/PhysRevD.66.074012}.

\bibitem[{\citenamefont{Cazaroto et~al.}(2013)\citenamefont{Cazaroto,
  Gon\ifmmode~\mbox{\c{c}}\else \c{c}\fi{}alves, and
  Navarra}}]{PhysRevD.88.034005}
\bibinfo{author}{\bibfnamefont{E.~R.} \bibnamefont{Cazaroto}},
  \bibinfo{author}{\bibfnamefont{V.~P.}
  \bibnamefont{Gon\ifmmode~\mbox{\c{c}}\else \c{c}\fi{}alves}},
  \bibnamefont{and} \bibinfo{author}{\bibfnamefont{F.~S.}
  \bibnamefont{Navarra}}, \bibinfo{journal}{Phys. Rev. D}
  \textbf{\bibinfo{volume}{88}}, \bibinfo{pages}{034005}
  (\bibinfo{year}{2013}),
  \urlprefix\url{https://link.aps.org/doi/10.1103/PhysRevD.88.034005}.

\bibitem[{\citenamefont{Maciu\l{}a and Szczurek}(2017)}]{PhysRevD.96.074013}
\bibinfo{author}{\bibfnamefont{R.}~\bibnamefont{Maciu\l{}a}} \bibnamefont{and}
  \bibinfo{author}{\bibfnamefont{A.}~\bibnamefont{Szczurek}},
  \bibinfo{journal}{Phys. Rev. D} \textbf{\bibinfo{volume}{96}},
  \bibinfo{pages}{074013} (\bibinfo{year}{2017}),
  \urlprefix\url{https://link.aps.org/doi/10.1103/PhysRevD.96.074013}.

\bibitem[{\citenamefont{Abazov et~al.}(2014)}]{Abazov:2014qba}
\bibinfo{author}{\bibfnamefont{V.~M.} \bibnamefont{Abazov}}
  \bibnamefont{et~al.} (\bibinfo{collaboration}{D0}), \bibinfo{journal}{Phys.
  Rev.} \textbf{\bibinfo{volume}{D90}}, \bibinfo{pages}{111101}
  (\bibinfo{year}{2014}), \eprint{1406.2380}.

\bibitem[{\citenamefont{Abazov et~al.}(2016)}]{Abazov:2015fbl}
\bibinfo{author}{\bibfnamefont{V.~M.} \bibnamefont{Abazov}}
  \bibnamefont{et~al.} (\bibinfo{collaboration}{D0}), \bibinfo{journal}{Phys.
  Rev. Lett.} \textbf{\bibinfo{volume}{116}}, \bibinfo{pages}{082002}
  (\bibinfo{year}{2016}), \eprint{1511.02428}.

\bibitem[{\citenamefont{Aaboud et~al.}(2017)}]{Aaboud:2016fzt}
\bibinfo{author}{\bibfnamefont{M.}~\bibnamefont{Aaboud}} \bibnamefont{et~al.}
  (\bibinfo{collaboration}{ATLAS}), \bibinfo{journal}{Eur. Phys. J.}
  \textbf{\bibinfo{volume}{C77}}, \bibinfo{pages}{76} (\bibinfo{year}{2017}),
  \eprint{1612.02950}.

\bibitem[{\citenamefont{Khachatryan et~al.}(2017)}]{Khachatryan:2016ydm}
\bibinfo{author}{\bibfnamefont{V.}~\bibnamefont{Khachatryan}}
  \bibnamefont{et~al.} (\bibinfo{collaboration}{CMS}), \bibinfo{journal}{JHEP}
  \textbf{\bibinfo{volume}{05}}, \bibinfo{pages}{013} (\bibinfo{year}{2017}),
  \eprint{1610.07095}.

\bibitem[{\citenamefont{Mäntysaari and Schenke}(2016)}]{Mantysaari:2016ykx}
\bibinfo{author}{\bibfnamefont{H.}~\bibnamefont{Mäntysaari}} \bibnamefont{and}
  \bibinfo{author}{\bibfnamefont{B.}~\bibnamefont{Schenke}},
  \bibinfo{journal}{Phys. Rev. Lett.} \textbf{\bibinfo{volume}{117}},
  \bibinfo{pages}{052301} (\bibinfo{year}{2016}), \eprint{1603.04349}.

\bibitem[{\citenamefont{Klein et~al.}(2002)\citenamefont{Klein, Nystrand, and
  Vogt}}]{Klein:2002wm}
\bibinfo{author}{\bibfnamefont{S.~R.} \bibnamefont{Klein}},
  \bibinfo{author}{\bibfnamefont{J.}~\bibnamefont{Nystrand}}, \bibnamefont{and}
  \bibinfo{author}{\bibfnamefont{R.}~\bibnamefont{Vogt}},
  \bibinfo{journal}{Phys. Rev.} \textbf{\bibinfo{volume}{C66}},
  \bibinfo{pages}{044906} (\bibinfo{year}{2002}), \eprint{hep-ph/0206220}.

\bibitem[{\citenamefont{Adeluyi and Nguyen}(2012)}]{Adeluyi:2012sw}
\bibinfo{author}{\bibfnamefont{A.}~\bibnamefont{Adeluyi}} \bibnamefont{and}
  \bibinfo{author}{\bibfnamefont{T.}~\bibnamefont{Nguyen}}
  (\bibinfo{year}{2012}), \eprint{1210.3327}.

\bibitem[{\citenamefont{von Weizsacker}(1934)}]{vonWeizsacker:1934nji}
\bibinfo{author}{\bibfnamefont{C.~F.} \bibnamefont{von Weizsacker}},
  \bibinfo{journal}{Z. Phys.} \textbf{\bibinfo{volume}{88}},
  \bibinfo{pages}{612} (\bibinfo{year}{1934}).

\bibitem[{\citenamefont{Williams}(1934)}]{Williams:1934ad}
\bibinfo{author}{\bibfnamefont{E.~J.} \bibnamefont{Williams}},
  \bibinfo{journal}{Phys. Rev.} \textbf{\bibinfo{volume}{45}},
  \bibinfo{pages}{729} (\bibinfo{year}{1934}).

\bibitem[{\citenamefont{K\l{}usek-Gawenda and
  Szczurek}(2010)}]{PhysRevC.82.014904}
\bibinfo{author}{\bibfnamefont{M.}~\bibnamefont{K\l{}usek-Gawenda}}
  \bibnamefont{and} \bibinfo{author}{\bibfnamefont{A.}~\bibnamefont{Szczurek}},
  \bibinfo{journal}{Phys. Rev. C} \textbf{\bibinfo{volume}{82}},
  \bibinfo{pages}{014904} (\bibinfo{year}{2010}),
  \urlprefix\url{https://link.aps.org/doi/10.1103/PhysRevC.82.014904}.

\bibitem[{\citenamefont{Frankfurt et~al.}(2011)\citenamefont{Frankfurt,
  Strikman, and Weiss}}]{Frankfurt:2010ea}
\bibinfo{author}{\bibfnamefont{L.}~\bibnamefont{Frankfurt}},
  \bibinfo{author}{\bibfnamefont{M.}~\bibnamefont{Strikman}}, \bibnamefont{and}
  \bibinfo{author}{\bibfnamefont{C.}~\bibnamefont{Weiss}},
  \bibinfo{journal}{Phys. Rev.} \textbf{\bibinfo{volume}{D83}},
  \bibinfo{pages}{054012} (\bibinfo{year}{2011}), \eprint{1009.2559}.

\bibitem[{\citenamefont{Dulat et~al.}(2016)\citenamefont{Dulat, Hou, Gao,
  Guzzi, Huston, Nadolsky, Pumplin, Schmidt, Stump, and Yuan}}]{Dulat:2015mca}
\bibinfo{author}{\bibfnamefont{S.}~\bibnamefont{Dulat}},
  \bibinfo{author}{\bibfnamefont{T.-J.} \bibnamefont{Hou}},
  \bibinfo{author}{\bibfnamefont{J.}~\bibnamefont{Gao}},
  \bibinfo{author}{\bibfnamefont{M.}~\bibnamefont{Guzzi}},
  \bibinfo{author}{\bibfnamefont{J.}~\bibnamefont{Huston}},
  \bibinfo{author}{\bibfnamefont{P.}~\bibnamefont{Nadolsky}},
  \bibinfo{author}{\bibfnamefont{J.}~\bibnamefont{Pumplin}},
  \bibinfo{author}{\bibfnamefont{C.}~\bibnamefont{Schmidt}},
  \bibinfo{author}{\bibfnamefont{D.}~\bibnamefont{Stump}}, \bibnamefont{and}
  \bibinfo{author}{\bibfnamefont{C.~P.} \bibnamefont{Yuan}},
  \bibinfo{journal}{Phys. Rev.} \textbf{\bibinfo{volume}{D93}},
  \bibinfo{pages}{033006} (\bibinfo{year}{2016}), \eprint{1506.07443}.

\bibitem[{\citenamefont{Aurenche et~al.}(2005)\citenamefont{Aurenche,
  Fontannaz, and Guillet}}]{Aurenche:2005da}
\bibinfo{author}{\bibfnamefont{P.}~\bibnamefont{Aurenche}},
  \bibinfo{author}{\bibfnamefont{M.}~\bibnamefont{Fontannaz}},
  \bibnamefont{and} \bibinfo{author}{\bibfnamefont{J.~P.}
  \bibnamefont{Guillet}}, \bibinfo{journal}{Eur. Phys. J.}
  \textbf{\bibinfo{volume}{C44}}, \bibinfo{pages}{395} (\bibinfo{year}{2005}),
  \eprint{hep-ph/0503259}.

\bibitem[{\citenamefont{Blok et~al.}(2012)\citenamefont{Blok, Dokshitser,
  Frankfurt, and Strikman}}]{Blok:2011bu}
\bibinfo{author}{\bibfnamefont{B.}~\bibnamefont{Blok}},
  \bibinfo{author}{\bibfnamefont{{\relax Yu}.}~\bibnamefont{Dokshitser}},
  \bibinfo{author}{\bibfnamefont{L.}~\bibnamefont{Frankfurt}},
  \bibnamefont{and} \bibinfo{author}{\bibfnamefont{M.}~\bibnamefont{Strikman}},
  \bibinfo{journal}{Eur. Phys. J.} \textbf{\bibinfo{volume}{C72}},
  \bibinfo{pages}{1963} (\bibinfo{year}{2012}), \eprint{1106.5533}.

\bibitem[{\citenamefont{Blok et~al.}(2014)\citenamefont{Blok, Dokshitzer,
  Frankfurt, and Strikman}}]{Blok:2013bpa}
\bibinfo{author}{\bibfnamefont{B.}~\bibnamefont{Blok}},
  \bibinfo{author}{\bibfnamefont{{\relax Yu}.}~\bibnamefont{Dokshitzer}},
  \bibinfo{author}{\bibfnamefont{L.}~\bibnamefont{Frankfurt}},
  \bibnamefont{and} \bibinfo{author}{\bibfnamefont{M.}~\bibnamefont{Strikman}},
  \bibinfo{journal}{Eur. Phys. J.} \textbf{\bibinfo{volume}{C74}},
  \bibinfo{pages}{2926} (\bibinfo{year}{2014}), \eprint{1306.3763}.

\bibitem[{\citenamefont{Chang et~al.}(2013{\natexlab{b}})\citenamefont{Chang,
  Manohar, and Waalewijn}}]{Chang:2012nw}
\bibinfo{author}{\bibfnamefont{H.-M.} \bibnamefont{Chang}},
  \bibinfo{author}{\bibfnamefont{A.~V.} \bibnamefont{Manohar}},
  \bibnamefont{and} \bibinfo{author}{\bibfnamefont{W.~J.}
  \bibnamefont{Waalewijn}}, \bibinfo{journal}{Phys. Rev.}
  \textbf{\bibinfo{volume}{D87}}, \bibinfo{pages}{034009}
  (\bibinfo{year}{2013}{\natexlab{b}}), \eprint{1211.3132}.

\end{thebibliography}
%------------------------

\end{document}